\documentclass[12pt]{article}
\usepackage{amsmath,amssymb,cite}

\numberwithin{equation}{section}

\begin{document}
\begin{flushright}
UCD-2001-02\\
\end{flushright}
\begin{center}
{\large{\bf PERTURBATIVE DYNAMICS ON THE FUZZY $S^2$ AND ${\mathbb
R}$P$^2$}}
\bigskip 

Sachindeo Vaidya\footnote{vaidya@dirac.ucdavis.edu}\\ 
{\it Department of Physics, \\
University of California, Davis, CA 95616,USA.} \\

\end{center}

\begin{abstract}
By considering scalar theories on the fuzzy sphere as matrix models,
we construct a renormalization scheme and calculate the one-loop
effective action. Because of UV-IR mixing, the two- and the four-point
correlators at low energy are not slowly varying functions of external
momenta. Interestingly, we also find that field theories on fuzzy
${\mathbb R}$P$^2$ avoid UV-IR mixing and hence are much more like
conventional field theories. We calculate the one-loop
$\beta$-function for the $O(N)$ theory on fuzzy ${\mathbb R}$P$^2$ at
large $N$ and show that in addition to the trivial one, it has a
nontrivial fixed point that is accessible in perturbation theory.
\end{abstract}

\section{Introduction}

Noncommutative theories have generated considerable interest since the
discovery that they arise in a certain limit of string theory
\cite{seiwitt}. They occur in the low energy descriptions of a class
of brane configurations that have a non-zero $B_{NS}$-field turned
on. Noncommutative manifolds also make their appearance in discussions
of brane configurations in external Ramond-Ramond (RR) fields
\cite{myers}.

Theories on noncommutative manifolds also offer a novel method of
discretization that is quite different from lattice discretization. In
this approach the manifold ${\cal M}$ is treated as a phase space and
is ``quantized''. If the manifold is compact, the total number of
states is finite and we end up with a matrix model. The continuum
physics corresponds to the limit in which the coordinates commute. 

Quantum properties of noncommutative theories are often very different
from the corresponding conventional ones, however. A striking
demonstration of this fact is the peculiar mixing between ultraviolet
(UV) and infrared (IR) degrees of freedom that occurs in quantum
noncommutative theories \cite{mrs}. More concretely, the correlation
functions show extreme sensitivity to low momenta because of this
mixing. In fact, it was shown by \cite{mrs} that the two-point
correlation function develops new poles and branch cuts at zero
momentum. This makes the low energy analysis considerably more subtle.

In this article, we will explore the quantum version of a class of
noncommutative theories in a controlled setting: scalar theories with
quartic interaction on a fuzzy $S^2$ and ${\mathbb R}$P$^2$. Because
these manifolds are compact, modes of the scalar field are quantized
and labelled by an angular momentum label $l$. Working on a compact
manifold also provides a natural infrared cutoff, which gives us a
better understanding about the nature of the IR singularities of the
correlation functions. Also, since theories on the fuzzy $S^2$ are
simply matrix models, we will be able to develop a scheme for doing
renormalization by integrating out ``shells'' of high momenta. Roughly
speaking, integrating out a single high (angular) momentum shell
corresponds to writing the theory in terms of a matrix of smaller
dimension. Rescaling corresponds to embedding this smaller matrix into
a matrix of original size that we started with. In the case of
ordinary theories, this procedure of integrating out a high momentum
shell allows us to write down the renormalization group (RG) equations
for the various couplings. We will show that as a consequence of UV-IR
mixing, the low-momentum correlation functions will behave very
differently for even and odd (angular) momenta. In other words, there
is no smooth way to approach zero momentum.

We also examine field theories on fuzzy ${\mathbb R}$P$^2$,
i.e. theories constructed from fields that take the same value at
antipodes of the two-sphere. Interestingly, we find that these
theories show no UV-IR mixing and behave much more like conventional
field theories. In particular, we will demonstrate that they have a
low energy effective Wilsonian description, which will allow us to
calculate the 1-loop $\beta$-function. For a single component scalar
theory, we will find that the non-trivial fixed point is very far away
from the zero mass Gaussian fixed point, and hence its existence
cannot be trusted. For $O(N)$ theories however, there is a nontrivial
fixed point at small coupling if $N$ is sufficiently large, and the
perturbative calculation is trustworthy.

The earliest investigations of the fuzzy sphere $S^2_F$ were by done
by \cite{madore}, followed by works of \cite{gkp1,gkp2}. Solitons and
monopoles in non-linear $\sigma$-models were studied by \cite{bbvy}
(see also \cite{gkp3}). Topological issues such as instantons,
$\theta$-term and derivation of the chiral anomaly on fuzzy $S^2$ were
discussed in \cite{balvai}. (For an alternate derivation of the chiral
anomaly, see \cite{presnajder}.) The continuum limit of the fuzzy
non-linear $\sigma$-model has been discussed in
\cite{bamaoc}. Interest in $S^2_F$ has also increased since Myers
showed that $D0$-branes in a constant $RR$ field arrange themselves in
the form of a fuzzy sphere \cite{myers}. There have also been
investigations by \cite{alresc} regarding open string versions of WZW
models which naturally lead to $S_F^2$. Gauge theories on $S_F^2$ have
also been studied by \cite{klimcik,watamura}. Continuum limits of
gauge theories on fuzzy sphere have also been discussed by
\cite{iktw}.

The organization of the paper is as follows: in section 2, we describe
scalar field theories on fuzzy $S^2$ and ${\mathbb R}$P$^2$, and make
notational explanations. The details of the renormalization group
procedure, namely decimation (i.e. integrating out a high energy
shell) and rescaling, are explained in section 3. In section 4, we
apply these techniques to quartic theories on fuzzy $S^2$ and
demonstrate UV-IR mixing. We also look at theories on ${\mathbb
R}$P$^2$ and derive the RG equations in section 5. The generalization
to $O(N)$ theories is easy, and we show that there is a nontrivial
fixed point at large $N$. Our conclusions are in section 6.

\section{Scalar theory on the fuzzy sphere}

Scalar theories on the fuzzy sphere were first discussed by
\cite{gkp1,gkp2}. We will review their work in a notation convenient
for our purposes.
 
The fuzzy sphere $S_F^2$ is described by three operators $X_a$
subject to the relations 
\begin{equation}
\sum_{a=1}^3 X_a X_a = R^2 {\bf 1}, \text{and} \quad [X_a, X_b] =
\frac{R i \epsilon_{abc}}{\sqrt{j(j+1)}}X_c.
\end{equation} 
The limit $j \rightarrow \infty$ reproduces the ordinary sphere
$\sum X_a X_a = R^2$. We will henceforth work with $R=1$.

Functions on an ordinary sphere may be written as
\begin{equation} 
f(x_a) = \sum(f_0 + f_a x_a + f_{ab}x_a x_b + \cdots) \equiv \sum_{l,m}
f_{lm} Y_{lm}(\theta, \phi).
\end{equation}
where $Y_{lm}(\theta, \phi)$ are the spherical harmonics. Under the
replacement $X_a \rightarrow (1/\sqrt{j(j+1)}) J_a$, $f$ is not a
function anymore, but a $(2j+1) \times (2j+1)$ matrix. The set of all
such matrices forms an algebra ${\cal A}_j$, which is the analog of
the algebra of functions on an ordinary sphere. In particular,
hermitian matrices are analogous to real functions. The scalar product
on ${\cal A}_j$ is defined by
\begin{equation} 
(f, g)_j = \frac{1}{2j+1} \text{Tr}(f^{\dagger} g), \quad f,g \in
{\cal A}_j.
\label{innerprod}
\end{equation} 
An arbitrary ``function'' (i.e. a matrix) on the fuzzy sphere can be
expanded in terms a special basis of matrices, namely the polarization
operators $T^{(j)}_{lm}$. These are constructed out of the angular
momentum operators $J_3, J_{\pm}=J_1 \pm i J_2$ and various powers
thereof (see for example \cite{vmkBook}). For example, $T^{(j)}_{00} =
\frac{1}{2j+1}{\bf 1}, T^{(j)}_{1m} =
-\sqrt{\frac{3}{2j(j+1)(2j+1)}} J_{\pm}$, etc. They are rank
$l$ irreducible tensors
\begin{equation}
[J_{\pm}, T^{(j)}_{lm}]= \sqrt{l(l+1)-m(m\pm 1)}T^{(j)}_{l,m \pm 1},
\quad [J_3, T^{(j)}_{lm}]= m T^{(j)}_{lm}, 
\end{equation} 
and satisfy the relations
\begin{equation} 
(T^{(j)}_{lm})^{\dagger} = (-1)^m T^{(j)}_{l,-m}, \quad \text{Tr}
[T^{(j)}_{lm} T^{(j)}_{l' m'}] = (-1)^m \delta_{l l'} \delta_{m+m', 0}. 
\end{equation} 
These allow us to expand any matrix $\Phi$ as
\begin{equation} 
\Phi = \sum^{2j}_{l=0}\sum^{l}_{m=-l} \phi_{lm} [(2j+1)^{1/2}T^{(j)}_{lm}]
\equiv \sum \phi_{lm} \Psi^{(j)}_{lm}.
\label{modeexp}
\end{equation}
The $\Psi^{(j)}_{lm}$'s form an orthonormal set with respect to the product
(\ref{innerprod}). The inner product of two arbitrary ``functions''
has the correct $j \rightarrow \infty$ limit.
 
If $\Phi$ is hermitian, then $\phi_{l, -m} = (-1)^{m}
\bar{\phi}_{lm}$. Thus for a given $l$, the total number of
independent real parameters is $2l+1$.

We will study actions for for $\Phi$ that are of the form
\begin{eqnarray} 
S &=&  \frac{1}{2j+1}\text{Tr} \left[-[J_i, \Phi] [J_i,\Phi] + \mu_j^2
\Phi^2 + \sum_n \frac{\lambda_j^{(n)}}{n!} \Phi^n \right], \nonumber \\   
&=&  \frac{1}{2j+1} \text{Tr}\left[\Phi[J_i, [J_i, \Phi]] + \mu_j^2
\Phi^2 + \sum_n \frac{\lambda_j^{(n)}}{n!} \Phi^n \right], \nonumber \\  
&\equiv& S_0 + S_{int}.
\label{scalaraction}
\end{eqnarray}	 
As $j \rightarrow \infty$, this action goes over to
\begin{equation} 
S = \frac{1}{4 \pi} \int d \Omega \left[-{\cal L}_i \Phi {\cal L}_i
\Phi + \mu^2 \Phi^2 + \sum_n \frac{\lambda^{(n)}}{n!}  \Phi^n
\right], \quad \text{where}\quad {\cal L}_i = -i \epsilon_{ijk}x_j \partial_k. 
\end{equation} 
This is the same as the action for a scalar field on the the sphere.

Using (\ref{modeexp}), we write the free action as
\begin{equation} 
S_0 = \left[\sum_{l=0}^{2j} \sum_{m=-l}^l |\phi_{lm}|^2 [l(l+1) +
\mu_j^2] \right],  
\end{equation}
and the quartic interaction term $\Phi^4$ as
\begin{eqnarray}
S_{int} &=& \frac{\lambda_j}{4!}\frac{1}{2j+1} \text{Tr}\Phi^4 \nonumber \\
&=& \frac{\lambda_j}{4!} (2j+1) \phi_{l_1 m_1}\phi_{l_2 m_2}\phi_{l_3
m_3}\phi_{l_4 m_4} \text{Tr}[\Psi^{(j)}_{l_1 m_1}\Psi^{(j)}_{l_2
m_2}\Psi^{(j)}_{l_3 m_3}\Psi^{(j)}_{l_4 m_4}] \nonumber \\ 
&\equiv&\phi_{l_1 m_1}\phi_{l_2 m_2}\phi_{l_3 m_3}\phi_{l_4
m_4}V(l_1,m_1;l_2,m_2;l_3,m_3;l_4,m_4;j).  
\end{eqnarray}
We will henceforth use the shorthand $V(1234;j)$ for the function
$V(l_1,m_1;l_2,m_2;l_3,m_3;l_4,m_4;j)$ from now on. It can be
conveniently written as
\begin{eqnarray} 	 
\lefteqn{V(1234;j) = \frac{\lambda_j}{4!} (2j+1)\prod_{i=1}^{4}(2l_i
+1)^{1/2} \times} \nonumber \\
&& \sum_{l,m}^{l=2j} \left\{ \begin{array}{ccc}
				l_1 & l_2 & l \\
				j   & j   & j
			  \end{array} \right\} \left\{ \begin{array}{ccc}
						l_3 & l_4 & l \\
						j   & j   & j
			  			\end{array} \right\}
(-1)^m C_{m_1 m_2 m}^{l_1\;\; l_2\;\; l} C_{m_3 m_4 -m}^{l_3\;\; l_4\;\; l}.
\label{quarticint}
\end{eqnarray}
The $C_{m_1 m_2 m}^{l_1\;\; l_2\;\; l}$ are the Clebsch-Gordan
coefficients and the objects with 6 entries within brace brackets are
the $6j$ symbols.

The partition function for the theory is
\begin{eqnarray} 
{\cal Z}_j &=& \int {\cal D}[\Phi]e^{-(S_0 + S_{int})} \quad \text{where} \\
{\cal D}[\Phi] &=& \prod_{lm}^{l=2j} \frac{d\bar{\phi}_{lm} d
\phi_{lm}}{2 \pi i}.
\end{eqnarray}
Correlation functions 
\begin{equation} 
\langle \phi_{l_1 m_1}\cdots \phi_{l_k m_k} \rangle = \int {\cal
D}[\Phi] \phi_{l_1 m_1}\cdots \phi_{l_k m_k}  e^{-(S_0 + S_{int})}
\end{equation} 
can be calculated using the standard procedure, by adding a source
term $\frac{1}{2j+1} \text{Tr} (J \Phi) = \sum_{lm} (\bar{J}_{lm}
\phi_{lm} + J_{lm} \bar{\phi}_{lm})$ to the action and then taking
derivatives with respect to $J_{lm}$. In particular, the two-point
correlation function or the propagator for the free action is
\begin{equation} 
\langle \phi_{l_1 m_1} \phi_{l_2 m_2}\rangle = \frac{\delta_{l_1 l_2}
\delta_{m_1 + m_2, 0}(-1)^{m_2}}{l(l+1) + \mu_j^2}.
\label{prop}
\end{equation} 

\subsection{Scalar theories on fuzzy ${\mathbb R}$P$^2$}
The manifold ${\mathbb R}$P$^2$ is obtained from $S^2$ by identifying
diametrically opposite points $\hat{r}$ and $-\hat{r}$. Functions on
${\mathbb R}$P$^2$ are simply a subset of the functions on $S^2$ that are
invariant under this identification. This means that only even $l$
values are allowed when the function is expanded in terms of spherical
harmonics, because $Y_{lm}(-\hat{r})=(-1)^l Y_{lm}(\hat{r})$. 

In the noncommutative case, we will be interested in fields that are
invariant under $\vec{J} \rightarrow -\vec{J}$. Again, since
$\Psi^{(j)}_{lm}(-\vec{J}) = (-1)^l \Psi^{(j)}_{lm}(\vec{J})$, we have
to restrict $l$ to even values. We stress that it is only in this
sense that we use the phrase `fuzzy ${\mathbb R}$P$^2$'.

We can now write the scalar action on the fuzzy ${\mathbb R}$P$^2$: it
is of the same form as (\ref{scalaraction}), with the additional
restriction that all the $l$'s be even, and hence $j$ an integer.

\subsection{Multicomponent Scalar field theories}
It is easy to generalize to $O(N)$ scalar field theories. There are
$N$ hermitian scalar fields $\Phi^\alpha, \alpha=1 \cdots N$. The
``linear'' $\sigma$-model on $S_F^2$ has the action
\begin{equation} 
S = \frac{1}{2j+1} \text{Tr}\left(- [J_i, \Phi^\alpha][J_i, \Phi^\alpha] +
\mu_j^2 \Phi \cdot \Phi + V(\Phi \cdot \Phi)\right)
\end{equation} 
where $\Phi \cdot \Phi = \sum_{\alpha=1}^N \Phi^\alpha
\Phi^\alpha$. We expand the fields $\Phi^\alpha$ as
\begin{equation} 
\Phi^\alpha = \sum_{l,m}^{l=2j} \phi^\alpha_{lm} \Psi^{(j)}_{lm}.
\end{equation} 
The free action can be evaluated to be
\begin{equation} 
S_{free} = \sum_{\alpha} \sum_{l,m} (l(l+1)+\mu_j^2)|\phi^\alpha_{lm}|^2  
\end{equation} 
and the propagator is
\begin{equation} 
\langle \phi^\alpha_{l_1 m_1} \phi^\beta_{l_2 m_2}\rangle =
\frac{\delta^{\alpha \beta} \delta_{l_1 l_2} \delta_{m_1 + m_2, 0}
(-1)^{m_2}}{l(l+1) + \mu_j^2}.  
\end{equation} 
If $V(\Phi \cdot \Phi) = \frac{\lambda_j}{4!}(\Phi \cdot \Phi)^2$,
then we write this quartic interaction as
\begin{equation}
S_{int} =  \sum_{\alpha,\beta} \sum_{l_i,m_i}\phi^\alpha_{l_1
m_1}\phi^\alpha_{l_2 m_2}\phi^\beta_{l_3 m_3}\phi^\beta_{l_4 m_4}
V(1234;j),  
\end{equation} 
where $V(1234;j)$ is the same as in (\ref{quarticint}).

For fuzzy ${\mathbb R}$P$^2$ the formulae are the same as above, but
with the restriction that all the $l$'s are even and $j$ is an
integer.

\section{Renormalization Procedure}

From our expansion for the matrix $\Phi$, we notice that the
``energy'' of the mode $\phi_{lm}$ increases as $l(l+1)$. In other
words, modes with large $l$ correspond to high energy
fluctuations. Our strategy from renormalization will be very simple
and very much in the spirit described by Wilson \cite{wilkog} (there
are many excellent reviews as well, see for
e.g. \cite{goldenfeld,shankar}). We start with a theory with some
large value of the cutoff $j$. The field $\Phi$ is described in terms
of $(2j+1) \times (2j+1)$ matrices, and the theory has modes going all
the way till $\phi_{2j,m}$. We separate the modes $\phi_{2j,m}$ as the
fast degrees of freedom and explicitly integrate out these modes in
the partition function, to get an effective action that depends only
on the modes $\phi_{00}, \cdots, \phi_{2j-1,-m}, \cdots,
\phi_{2j-1,m}$. For large $j$, the effective action should be
describable in terms of the field $\Phi_{slow} = \sum_{lm}^{l=2j-1}
\phi_{lm}\Psi^{(j-\frac{1}{2})}_{lm}$, i.e. in terms of $2j \times 2j$
matrices. We now ``rescale'' the field $\Phi_{slow}$ by embedding the
$2j \times 2j$ matrix in a $(2j+1) \times (2j+1)$ matrix. This
rescaling redefines
\begin{eqnarray} 
\phi_{lm} &\rightarrow& \phi'_{l' m'} \quad l'=0 \cdots 2j, \\
\mu^2_j &\rightarrow& {\mu'}^2_j, \\
\lambda_j &\rightarrow& \lambda'_j,
\end{eqnarray} 
and so on. Comparing the rescaled action to the original one gives us
the renormalization group equations for $\mu^2_j$ and
$\lambda_j$. Since rescaling corresponds to $(2j-1) \rightarrow 2j$,
the rate of change of the various couplings is given by the equations
\begin{eqnarray} 
\frac{d \mu_j^2}{dt} \equiv 2j ({\mu'}_j^2 - \mu_j^2) &=& M(\mu_j^2,
\lambda_j), \\
\frac{d \lambda_j}{dt} \equiv 2j ({\lambda'}_j - \lambda_j) &=&
G(\mu_j^2, \lambda_j)
\end{eqnarray}
where $dt$ is given by $(2j-1)/2j = e^{dt}$. In particular the zeroes
of $G(\mu_j^2, \lambda_j)$ tells us the fixed points of the
$\beta$-function.

Let us first understand the scaling of the field $\Phi$. Recall that
the free massless action is
\begin{equation} 
S = \frac{1}{2j+1} \text{Tr}( \Phi [J_i,[J_i, \Phi]]) =
\sum_{l=0}^{2j} \sum_{m=-j}^{j} |\phi_{lm}|^2[l(l+1)].
\end{equation} 
The action for the slow variables is simply 
\begin{equation}	
S_{slow}=\sum_{l=0}^{2j-1} \sum_{m=-j+1/2}^{j-1/2}
|\phi_{lm}|^2[l(l+1)] 
\end{equation} 
which we rewrite as 
\begin{equation} 
S_{slow} = \frac{1}{2j} \text{Tr} (\Phi_{slow}[J_i,[J_i, \Phi_{slow}]]),
\end{equation}
where
\begin{equation} 
\Phi_{slow} = \sum_{l=0}^{2j-1} \phi_{lm} \Psi^{(j-\frac{1}{2})}_{lm},
\end{equation}	
the $\Psi^{(j-\frac{1}{2})}_{lm}$ being $2j \times 2j$ matrices. These
$\Psi^{(j-\frac{1}{2})}_{lm}$ are simply the orthonormal basis in
terms of which we can expand any $2j \times 2j$ matrix. We embed each
$\Psi^{(j-\frac{1}{2})}_{lm}$ in a $(2j+1) \times (2j+1)$ matrix as
\begin{equation} 
\left[ \begin{array}{cc}
	    \Psi^{(j-\frac{1}{2})}_{lm} & 0 \\
		0     &  0
	\end{array}
\right] \equiv {\Psi'}^{(j)}_{lm}
\end{equation}     
allowing us to write
\begin{equation} 
\Phi' = \sum_{l=0}^{2j-1} \phi_{lm} {\Psi'}^{(j)}_{lm}.
\end{equation}
This $\Phi'$ is a $(2j+1)\times(2j+1)$ matrix, and hence can be
re-expressed using the $\Psi^{(j)}_{lm}$'s:
\begin{equation} 
{\Phi'}^{(j)} = \sum_{l=0}^{2j} \phi'_{lm} \Psi^{(j)}_{lm}.
\end{equation} 
This is the rescaling operation in our RG procedure. It gives us the
relation between the $\phi_{lm}$'s and the $\phi'_{lm}$'s:
\begin{equation} 
\phi'_{l' m'} = \sum_{l,m}^{l=2j-1} \left[\phi_{lm} \sqrt{\frac{(2l+1)(2l'
+1)}{2j(2j+1)}} \left(\sum_{\nu, \nu'}C_{\;\;\nu \;\;\;\; m \;\;\;
\nu'}^{j-\frac{1}{2}\;\; l\;\; j-\frac{1}{2}} C_{\nu+\frac{1}{2} 
\;\;m' \;\;\nu' +\frac{1}{2}}^{\;\;\; j \quad l' \quad j} \right)\right].
\end{equation} 
Thus
\begin{equation} 
S_{slow} = \frac{1}{2j} \text{Tr}(\Phi' [J_i,[J_i, \Phi']]) =
\frac{2j+1}{2j} \sum_{l=0}^{2j} \sum_m |\phi'_{lm}|^2 [l(l+1)]
\end{equation} 
The actions $S$ and $S_{slow}$ can now be identified. In order that
the kinetic energy term look the same in the original and the rescaled
action, the $\phi$'s must be scaled as
\begin{equation} 
\phi_{lm} = \left(\frac{2j+1}{2j}\right)^{1/2} \phi'_{l'm'} \delta_{l
l'} \delta_{m m'} = \left(\frac{2j+1}{2j}\right)^{1/2} \phi'_{lm}
\label{wfscaling}
\end{equation}
for $l << 2j$. 
Similar considerations tell us that the mass $\mu_j^2$ scales as:
\begin{equation}
{\mu'}_j^2 = \frac{2j+1}{2j}\mu_j^2.
\end{equation} 

Now that we know how to do the rescaling, it is straightforward to
write down the effective action. We separate $S$ into $S_{slow}$ and
$S_{fast}$:
\begin{eqnarray}  
\lefteqn{{\cal Z} = \int {\cal D}[\Phi]e^{-(S_0 + S_{int})} = } \nonumber \\
&& \int {\cal D}[\Phi_{slow}] e^{-(S_0 (slow) + S_{int} (slow))} \int {\cal
D}[\Phi_{fast}] e^{-(S_0 (fast) + \delta S [\Phi_{slow}, \Phi_{fast}])}.   
\end{eqnarray}  
We explicitly do the integral over the fast variables $\Phi_{fast} =
\sum \phi_{2j, m}\Psi^{(j)}_{2j,m}$:
\begin{equation}
\int {\cal D}[\phi_{fast}]e^{-(S_0(fast) + \delta S [\Phi_{slow},
\Phi_{fast}])} \equiv \langle e^{- \delta S[\Phi_{slow}, \Phi_{fast}]}
\rangle_f = e^{-S'_{slow}}
\label{fastint}
\end{equation}
where $\langle {\cal O}\rangle_f$ stands for the expectation value of
${\cal O}$ over the fast degrees of freedom.

Finally, we rescale the variables so that the two actions look
similar:
\begin{equation} 
\int {\cal D}[\Phi_{slow}] e^{-(S_0 (slow) + S_{int} (slow) + S'_{slow})}
= \int {\cal D}[\Phi']e^{-(S_0 [\Phi'] + S_{int}[\Phi'])}.
\end{equation} 
A comparison of the coupling constants of the original theory and the
rescaled theory gives us the RG equations. 

For fields on fuzzy ${\mathbb R}$P$^2$, the analysis is almost
identical. Since $j$ takes only integer values, integrating out the
high energy modes $\phi_{2j,m}$ leaves us with the modes $\phi_{00},
\cdots, \phi_{2j-2,m}$, which are then reassembled into a $(2j-2)
\times (2j-2)$ matrix. Rescaling corresponds to $(2j-2) \rightarrow
2j$. For $l<<2j$ we find that $\phi_{lm}$ and $\mu_j^2$ must scale as
\begin{eqnarray} 
\phi_{lm} &=& \left(\frac{2j+2}{2j}\right)^{1/2} \phi'_{lm},
\label{wfscalingRP} \\
{\mu'}_j^2 &=& \frac{2j+2}{2j}\mu_j^2.
\end{eqnarray} 

\section{Quartic Theories}

We are now ready to apply the RG technique described above to quartic
interactions. The action cannot be evaluated exactly, and we will
resort to perturbation theory.

We separate the interaction piece $S_{int}$ of the full action into
slow and fast parts:
\begin{eqnarray}   
S_{int} &=& \phi_1 \phi_2 \phi_3 \phi_4 V(1234;j)  \nonumber \\
&=& S^{slow}_{int} + 4 \phi_1 \phi_2 \phi_3 \phi_{f_4} V(123f_4) + 4
\phi_1 \phi_2 \phi_{f_3} \phi_{f_4} V(12f_3 f_4)  \nonumber \\ 
&+& 2 \phi_1 \phi_{f_2} \phi_3 \phi_{f_4} V(1 f_2 3 f_4)+  4 \phi_1
\phi_{f_2} \phi_{f_3} \phi_{f_4} V(1 f_2 f_3 f_4) + S^{fast}_{int}
\nonumber \\ 
&\equiv&S^{slow}_{int} + \delta S_1 + \delta S_2 + \delta S_3
+S^{fast}_{int} = S^{slow}_{int} + \delta S.   
\end{eqnarray}
We have used the shorthand 
\begin{eqnarray} 
\phi_1 &=& \phi_{l_1 m_1}, \quad \phi_{f_1} = \phi_{2j,m_1}, \quad
\text{etc}\nonumber \\ 
V(123f_4) &=& V(l_1,m_1;l_2,m_2;l_3,m_3;2j,m_4;j) \quad \text{etc}. \nonumber 
\end{eqnarray} 
Having separated the action into slow and fast parts, we proceed to
evaluate (\ref{fastint}) using perturbation theory. In order to do so,
we make use of the cumulant expansion:
\begin{equation} 
\left \langle e^{-\delta S} \right \rangle_f = \exp{\left(-\langle \delta S
\rangle_f + \frac{\langle \delta S^2 \rangle_f - \langle \delta S
\rangle_f^2}{2} + \cdots \right)}.
\label{cumulant}
\end{equation}
The first term of the exponent is 
\begin{eqnarray}     	
\langle \delta S \rangle_f &=& \langle S^{fast}_{int}\rangle_f + 4
\phi_1 \phi_2 \phi_3 V(123f_4)\langle \phi_{f_4} \rangle_f + 4
\phi_1V(1 f_2 f_3 f_4) \langle \phi_{f_2} \phi_{f_3} \phi_{f_4}
\rangle_f  \nonumber \\ 
&+& 4 \phi_1 \phi_2 V(12f_3 f_4) \langle \phi_{f_3} \phi_{f_4} \rangle_f +
2 \phi_1 \phi_3 V(1 f_2 3 f_4) \langle \phi_{f_2} \phi_{f_4} \rangle_f. 
\end{eqnarray} 
The second and the third terms are zero, while the first term is an
irrelevant constant. The fourth term involves summing over $m_3$ and
$m_4$, each sum going over the range $-2j$ to $2j$. Similarly the last
term involves summing over $m_2$ and $m_4$. To calculate these sums,
we will make extensive use of identities for summing $3j$- and $6j$
symbols found in \cite{vmkBook}. Using (\ref{prop}), we can write this
term as
\begin{equation} 
\langle \delta S \rangle_f = |\phi_{lm}|^2 \frac{\lambda_j}{4!}
\left[\frac{4(4j+1)}{2j(2j+1) + \mu_j^2} + (-1)^l
\frac{2(2j+1)(4j+1)}{2j(2j+1)+\mu_j^2} \left\{ \begin{array}{ccc}
				                 j & j & l \\
				                 j & j & 2j
			                      \end{array} \right\}
\right].
\end{equation} 
The scaling relation for $\phi$'s (\ref{wfscaling}) gives us the
renormalized two-point vertex:
\begin{equation}
G^{-1}(\phi \phi') = \langle \phi_{lm} \phi_{l' m'} \rangle^{-1} =
\delta_{l l'} \delta_{m + m',0} (-1)^m [l(l+1) + {\mu'}^2_j] 
\end{equation} 
where
\begin{equation} 
{\mu'}^2 = \left(\frac{2j+1}{2j}\right)\left[\mu_j^2 + \frac{2 \lambda_j}{4!}
A(2j) (2+ (-1)^l B(l,2j)) \right],
\label{massRG}
\end{equation} 
and
\begin{equation} 
A(2j) = \frac{2(2j)+1}{2j(2j+1)+\mu_j^2}, \quad B(l,2j)= \frac{(2j+1)!
(2j)!}{(2j-l)! (2j+l+1)!}.
\label{defAB}
\end{equation}
The term in (\ref{massRG}) with $(-1)^l$ is due to noncommutative
effects, and is not present in ordinary theories. 

To calculate the one-loop contribution to the four-point function, we
need the next term of the cumulant expansion (\ref{cumulant}). This
automatically counts only the connected diagrams. The contribution
to the four-point function comes only from $\delta S_2$. We find that 
\begin{eqnarray}       	  
\langle \delta S_2^2 \rangle &=& 16 \phi_1 \phi_2 \phi_{1'} \phi_{2'}
V(12f_3 f_4)  V(1' 2' f_{3'} f_{4'}) \langle \phi_{f_3} \phi_{f_4}
\phi_{f_{3'}} \phi_{f_{4'}} \rangle_f \nonumber \\ 
&+& 16 \phi_1 \phi_2 \phi_{1'} \phi_{3'} V(12 f_3 f_4)V(1' f_{2'} 3'
f_{4'}) \langle \phi_{f_3} \phi_{f_4}\phi_{f_{2'}} \phi_{f_{4'}}
\rangle_f \nonumber \\ 
&+& 4 \phi_1 \phi_3 \phi_{1'} \phi_{3'} V(1 f_2 3 f_4) V(1' f_{2'} 3'
f_{4'})\langle \phi_{f_2} \phi_{f_4} \phi_{f_{2'}} \phi_{f_{4'}}.  
\end{eqnarray} 

The three terms can be evaluated using various Clebsch-Gordan
identities \cite{vmkBook} to give the following:
\begin{eqnarray} 
\lefteqn{\frac{1}{2}\left(\langle \delta S^2 \rangle - \langle \delta S
\rangle^2 \right) = \frac{\lambda_j^2}{(4!)^2} \sum_{l_i,m_i} \phi_1
\phi_2 \phi_3 \phi_4 A^2(2j)(2j+1)^2 \prod_{i=1}^{4}(2l_i +1)^{1/2}
\times} \nonumber \\
&&\left[ \sum_{l,m} F(l_i,m_i;l)(\Gamma_1(l_i;l,2j) +
\Gamma_2(l_i;l,2j) + \Gamma_3(l_i;l,2j)\right], \\
&\equiv&\sum_{l_i,m_i} \phi_1 \phi_2 \phi_3 \phi_4 \Gamma^{(1-loop)}_4
(l_i,m_i;j)   
\end{eqnarray}  
where $A(2j)$ is from (\ref{defAB}) and
\begin{eqnarray} 
F(l_i,m_i;l) &=& (-1)^m C_{m_1 m_2 m}^{l_1\;\; l_2\;\; l} C_{m_3 m_4
-m}^{l_3\;\;l_4\;\; l}, \\
\Gamma_1(l_i;l,2j) &=& 8(1+(-1)^l) \times \nonumber \\
&& \times \left\{ \begin{array}{ccc}
	             l_1 & l_2 & l \\
	             j   & j   & j
		  \end{array} \right\} 
\left\{ \begin{array}{ccc} 
             l_3 & l_4 & l \\
	     j   & j   & j
	\end{array} \right\} \left\{ \begin{array}{ccc}
				2j & 2j & l \\
				j   & j   & j
			  \end{array} \right\}^2, \\
\Gamma_2(l_i;l,2j) &=& 8 (-1)^{2j+l_4} (1+(-1)^l) \times \nonumber \\
&& \times \left\{ \begin{array}{ccc}
            l_1 & l_2 & l \\
	    j   & j   & j
	\end{array} \right\}\left\{ \begin{array}{ccc}
				       2j & 2j & l \\
				      j   & j   & j
			  	     \end{array} \right\} 
\left\{ \begin{array}{ccc}
	    l_3 & l_4 & l \\
	    j   & j   & 2j\\
            j   & j   & 2j
        \end{array}\right\}, \\
\Gamma_3(l_i;l,2j) &=&2[(-1)^{l_2 + l_4} + (-1)^{l_3 + l_4}](-1)^l
\times \nonumber \\
&& \times \left\{ \begin{array}{ccc}
           l_1 & l_2 & l \\
	   j   & j   & 2j\\
           j   & j   & 2j
	 \end{array}\right\} \left\{ \begin{array}{ccc}
				        l_3 & l_4 & l \\
					j   & j   & 2j\\
                                        j   & j   & 2j
			  	      \end{array} \right\},
\end{eqnarray}
where the objects with 9 entries within brace brackets are the
$9j$-symbols. This tells us the renormalized four-point function and
hence the RG equation for the quartic coupling:
\begin{equation} 
V'(1234;j)=\left(\frac{2j+1}{2j}\right)^2 \left[V(1234;j) -
\Gamma^{(1-loop)}_4 (l_i,m_i;j) \right]  
\label{couplingRG}
\end{equation}

\subsection{UV-IR mixing}
Let us recall how the $\beta$-function is calculated for ordinary
scalar theories. Integrating out the fast modes gives equations
analogous to (\ref{massRG},\ref{couplingRG}). These renormalized
$n$-point correlation functions are slowly varying functions of
external momenta $l_i$, at least for small momenta: for example at
large $j$, the 2-point function with $l_i=0$ differs only slightly
from that at $l_i=1$. If the momenta take continuous values, this
translates to saying that the correlation functions are analytic
functions of external momenta in the neighborhood of zero external
momentum. This allows us to derive a difference (or differential)
equation for the mass and the coupling constant.

The situation in our case is clearly different. From either
(\ref{massRG}) or (\ref{couplingRG}), it is obvious that the
correlation functions are not slowly varying functions of $l_i$ for
small values of the external momenta $l_i$: the relative factors of
$(-1)^{l_i}$ make the values of the correlation functions change
abruptly as one moves from $l_i$ to $l_i +1$. This is a clear
signature of UV-IR mixing: integrating out a high energy mode has a
violent effect on the low energy properties of the correlation
functions, and traditional Wilsonian RG cannot be implemented.

\section{RG equations on fuzzy ${\mathbb R}$P$^2$}

As explained before, scalar fields on ${\mathbb R}$P$^2$ come only with
even values of $l$. There are no factors of $(-1)^{l_i}$ in the two-
and four-point correlation functions, and so they are indeed slowly
varying functions of external momenta. They can thus be thought of as
coming from a low-energy Wilsonian action, and can be used to write
the RG equations for the mass and the coupling constant.

If we define the square of the mass and the coupling constant as the
two- and four-point vertices respectively at zero external momentum,
we get the following RG equations:
\begin{eqnarray} 
j({\mu'}_j^2 - \mu_j^2) &=& \mu^2 + \frac{6\lambda_j}{4!} [j A(2j)] +
O(1/j), \\
\beta(\lambda_j) \equiv j (\lambda'_j - \lambda_j) &=& 2 \lambda_j +
\frac{\lambda_j}{j} - \frac{4 \lambda_j^2}{4!} \frac{18
j(4j+1)}{[2j(2j+1)+\mu_j^2]^2}. 
\end{eqnarray}
Let $\epsilon=1/j$. For $\epsilon$ small, these equations can be
written as
\begin{eqnarray}  
\frac{{\mu'}_j^2 - \mu_j^2}{\epsilon}&=& \mu_j^2 + \frac{\lambda_j}{4}, \\
\beta(\lambda_j) &=& (2 + \epsilon) \lambda_j - \frac{3 \lambda_j^2
\epsilon^2}{4} + O(\epsilon^3).  
\end{eqnarray} 
The interpretation of the $\beta$-function equation is standard: the
first term tells us that the coupling increases until nonlinear
effects (described by the second term) kick in. There exists a
critical value $\lambda_c$ at which the increase in the coupling due
to rescaling is compensated by the decrease due to the nonlinear
effect. However this non-trivial fixed point occurs at $\lambda_c =
4(2+ \epsilon)/3\epsilon^2$ i.e. at large $\lambda$, since $\epsilon =
1/j$ is small. This fixed point was deduced using perturbation theory,
and so its existence cannot be trusted. In order to know if there is
really a non-trivial fixed point, one would need to know the
$\beta$-function to all orders in $\lambda$. We do not know how to do
this, but can estimate the two-loop contribution. At large $j$,
\begin{equation} 
\beta(\lambda_j) = (2+\epsilon) \lambda_j - \frac{3 \lambda_j^2
\epsilon^2}{4} + C \lambda_j^3 \epsilon^5, 
\end{equation} 
where $C$ is a constant of order 1. Although its numerical value
changes, the nontrivial fixed point continues to exist. However, as
mentioned before, in order to really trust the new fixed point, we
will need $\beta(\lambda_j)$ to all orders in $\lambda_j$.

\subsection{$O(N)$ theories on ${\mathbb R}$P$^2$} 
The large $N$ limit of $O(N)$ theories solves the problem we faced
regarding the fixed point of the $\beta$-function. The RG equations
for the mass and coupling constant are
\begin{eqnarray}
{\mu'}_j^2 &=& \frac{2j+2}{2j}\left[\mu_j^2 + \frac{(2N+2)\lambda_j}{4!}
\frac{4j+1}{2j(2j+1) + \mu_j^2} \right], \label{NRGmass} \\
\lambda_j' &=& \left(\frac{2j +2}{2j}\right)^2 \left[\lambda_j -
\frac{(8N+64)\lambda^2}{4!} \frac{4j+1}{[2j(2j+1)+\mu_j^2]^2}\right]. 
\label{NRGeqn}
\end{eqnarray}  
From (\ref{NRGeqn}, \ref{NRGmass}), it is easy to see that the
non-trivial fixed point is at
\begin{eqnarray}
\mu_c^2 &=& - \left(\frac{\frac{1}{2} \frac{N+1}{N+8} \frac{j(2j+1)^2}{j+1}}{1
+ \frac{1}{4} \frac{N+1}{N+8} \frac{2j+1}{j+1}}\right),  \\
\lambda_c &=& \frac{3}{N+8} \left(\frac{2j+1}{4j+1}\right)
\left(\frac{2j(2j+1) + \mu^2_c}{j+1}\right)^2, \\
&=& \frac{3}{N+8} \left(\frac{4j^2(2j+1)^3}{(j+1)^2 (4j+1)}\right)
\left(\frac{1}{1 + \frac{1}{4} \frac{N+1}{N+8}
\frac{2j+1}{j+1}} \right)^2. 
\end{eqnarray}   
Thus if
$N+8>>32j^2 /3$, the new fixed point occurs at small value of $\lambda_j$,
and its deduction is consistent with perturbation theory.

\section{Conclusion and Outlook}

We have done a careful perturbative analysis of quartic theories on
noncommutative $S^2$ and ${\mathbb R}$P$^2$. Although the classical
theory on the noncommutative $S^2$ is the same as the corresponding
theory on the ordinary $S^2$ in the limit of large $j$, the quantum
theories are very different. Even in the limit of large $j$, the low
energy behavior of correlation functions of the quantum theory on
noncommutative $S^2$ shows a mixing between UV and IR that is
characteristic of noncommutative theories.

Surprisingly, we also find that if we restrict ourselves field
theories on noncommutative ${\mathbb R}$P$^2$, we can avoid the
effects of UV-IR mixing. In the large $j$-limit, the quartic theory on
${\mathbb R}$P$^2$ flows away from the zero mass Gaussian fixed point
in the infrared. For a single component scalar theory, it is difficult
to find the new fixed point in perturbation theory. The situation for
$O(N)$ theories is much better. For $N$ sufficiently large, these
theories have a fixed point that is ``close'' to $\lambda=0$, and
hence can be trusted in perturbation theory.

There are several questions that may be studied in light of our
results. One can try to look for situations in supergravity/string
theory that correspond to branes distributed on a fuzzy ${\mathbb
R}$P$^2$ instead of a fuzzy $S^2$. If such configurations exist, our
results could have significant implications for their low energy
dynamics.

Compactification scenarios that use fuzzy torus as extra dimensions
have been suggested recently \cite{gomewi}. It may illuminating to
understand the implications of compactifying on the fuzzy sphere in
light of the non-trivial UV-IR mixing.

Our approach to quantum theories on fuzzy manifolds in this article
relies on such theories being expressible as models of
finite-dimensional matrices. In particular, the realization of the
sphere $S^2$ as a coadjoint orbit $SU(2)/U(1)$ allows us to make
extensive use of $SU(2)$ representation theory. Using the coadjoint
orbit method, one can also construct fuzzy versions of ${\mathbb
C}$P$^2$ and $\frac{SU(3)}{U(1) \times U(1)}$, which in the continuum
limit correspond to four- and six-dimensional manifolds respectively
\cite{trivai}. It would be interesting to use techniques analogous to
the ones in this article to study explicitly the nature of UV-IR
mixing on these fuzzy manifolds, and whether UV-IR mixing can be
avoided by imposing restrictions on the modes as we found for the case
of fuzzy ${\mathbb R}$P$^2$.

{\bf Acknowledgments:} It is a pleasure to thank A. P. Balachandran,
Sumit Das, Steve Carlip, Doug McKay, Jim Wells and especially Denjoe
O'Connor for discussions.  This work was supported in part by
Department of Energy grant DE-FG03-91ER40674.

\bibliographystyle{unsrt}

\bibliography{rg}

\begin{thebibliography}{10}

\bibitem{seiwitt}
Nathan Seiberg and Edward Witten.
\newblock {\em JHEP}, 09:032, 1999.
\newblock {\tt hep-th/9908142}.

\bibitem{myers}
R.~C. Myers.
\newblock {\em JHEP}, 12:022, 1999.
\newblock {\tt hep-th/9910053}.

\bibitem{mrs}
S.~Minwalla, M.~Van Raamsdonk and N.~Seiberg.
\newblock Noncommutative perturbative dynamics.
\newblock {\tt hep-th/9912072}.

\bibitem{madore}
J.~Madore.
\newblock {\em Class. Quant. Grav.}, 9:69--88, 1992.

\bibitem{gkp1}
H.~Grosse, C.~Klimcik and P.~Presnajder.
\newblock {\em Int. J. Theor. Phys.}, 35:231--244, 1996.
\newblock {\tt hep-th/9505175}.

\bibitem{gkp2}
H.~Grosse, C.~Klimcik and P.~Presnajder.
\newblock {\em Commun. Math. Phys.}, 185:155--175, 1997.
\newblock {\tt hep-th/9507074}.

\bibitem{bbvy}
S.~Baez, A.~P. Balachandran, S.~Vaidya and B.~Ydri.
\newblock {\em Commun. Math. Phys.}, 208:787--798, 2000.
\newblock {\tt hep-th/9811169}.

\bibitem{gkp3}
H.~Grosse, C.~Klimcik and P.~Presnajder.
\newblock {\em Commun. Math. Phys.}, 178:507--526, 1996.
\newblock {\tt hep-th/9510083}.

\bibitem{balvai}
A.~P. Balachandran and S.~Vaidya.
\newblock {\em Int. J. Mod. Phys.}, A16:17--40, 2001.
\newblock {\tt hep-th/9910129}.

\bibitem{presnajder}
P.~Presnajder.
\newblock {\em J. Math. Phys.}, 41:2789--2804, 2000.
\newblock {\tt hep-th/9912050}.

\bibitem{bamaoc}
A.~P. Balachandran, X.~Martin and D.~O'Connor.
\newblock Fuzzy actions and their continuum limits.
\newblock {\tt hep-th/0007030}.

\bibitem{alresc}
A.~Yu. Alekseev, A.~Recknagel and V.~Schomerus.
\newblock {\em JHEP}, 09:023, 1999.
\newblock {\tt hep-th/9908040}.

\bibitem{klimcik}
C.~Klimcik.
\newblock {\em Commun. Math. Phys.}, 199:257--279, 1998.
\newblock {\tt hep-th/9710153}.

\bibitem{watamura}
U.~Carow-Watamura and S.~Watamura.
\newblock {\em Commun. Math. Phys.}, 212:395--413, 2000.
\newblock {\tt hep-th/9801195}.

\bibitem{iktw}
S.~Iso, Y.~Kimura, K.~Tanaka and K.~Wakatsuki.
\newblock Noncommutative gauge theory on fuzzy sphere from matrix model.
\newblock {\tt hep-th/0101102}.

\bibitem{vmkBook}
D.~A. Varshalovich, A.~N. Moskalev and V.~K. Khersonsky.
\newblock {\em Quantum Theory of Angular Momentum}.
\newblock World Scientific, New Jersey, 1998.

\bibitem{wilkog}
K.~G. Wilson and J.~Kogut.
\newblock {\em Phys. Rept.}, 12:75--200, 1974.

\bibitem{goldenfeld}
N.~Goldenfeld.
\newblock {\em Lectures on Phase Transitions and the Renormalization Group}.
\newblock Perseus Books, Reading, Mass., 1992.

\bibitem{shankar}
R.~Shankar.
\newblock {\em Rev. Mod. Phys.}, 66:129--192, 1994.
\newblock {\tt cond-mat/9307009}.

\bibitem{gomewi}
J.~Gomis, T.~Mehen and M.~B. Wise.
\newblock {\em JHEP}, 08:029, 2000.
\newblock {\tt hep-th/0006160}.

\bibitem{trivai}
S.~P. Trivedi and S.~Vaidya.
\newblock {\em JHEP}, 09:041, 2000.
\newblock {\tt hep-th/0007011}.

\end{thebibliography}

\end{document}